\documentstyle[preprint,aps,tighten,floats,epsf]{revtex}
\preprint{ENSLAPP-A-645/97, SPhT/97-051}
\begin{document}
\draft
\title{CRITICAL EXPONENTS OF THE PURE AND RANDOM-FIELD ISING MODELS}
\author{Thierry Jolic\oe ur${}^1$ and
Jean-Claude Le Guillou$^2$\cite{IUF}\thanks{e-mail:
thierry@spht.saclay.cea.fr, leguil@lapp.in2p3.fr}}
\address{${}^1$Service de Physique Th\'eorique,
CE Saclay, F-91191 Gif-sur-Yvette, France}
\address{${}^2$Laboratoire de Physique Th\'eorique 
ENSLAPP,\cite{lab}
B.P. 110, F-74941 Annecy-le-Vieux, France}
\date{15 May 1997}
\maketitle
\begin{abstract}
We show that current estimates of the critical exponents of the
three-dimensional random-field Ising model are in agreement with the
exponents of the pure Ising system in dimension $3 - \theta $ where 
$\theta $ is the exponent that governs the hyperscaling violation
 in the random case. 
\end{abstract}

The phase transition of Ising systems in a random quenched field is
the subject of ongoing research\cite{Reviews}. It is known that in
the three-dimensional case, there is a low-temperature ordered phase.
This was originally suggested by a heuristic argument due to Imry 
and
Ma\cite{Ima} and has been shown rigorously\cite{Chalk,FFS,Imbrie}. In
the presence of random fields, these arguments show in fact that the
lower critical dimension is $d=2$ in the Ising case. The Hamiltonian
of such systems is given by:
\begin{equation}
H=-\sum_{\langle ij\rangle} S_i S_j -\sum_i h_i S_i
\label{RFI}
\end{equation}
where the Ising spins $S_i=\pm 1$ are on the sites of a $
d$-dimensional cubic lattice and interact only between
nearest-neighbors. The random fields $h_i$ have a quenched
probability distribution $\overline h_i =0$ and $\overline{h_i h_j}=
{\overline h}\delta_{ij}$ where the overbar stands for average over
the disorder. There is accumulating evidence for a second-order phase
transition in three dimensions from Monte-Carlo
studies\cite{OH,RIY,RI}, real-space renormalization group
calculations\cite{Dayan,RSRG,FH} as well as series expansions\cite{series}.

The standard perturbative approach to critical phenomena leads to an
upper critical dimension\cite{Gri} $d=6$ for the random field problem
and, below this dimension, the critical exponents are those of the
pure Ising system in dimension $d-2$ within the $\epsilon=d-6$
expansion. In fact this dimensional reduction holds to all orders in
perturbation theory\cite{Gri,AIM,APY,PaSo} in $\epsilon$. In view
of the lower critical dimension $d=2$, this cannot be the whole
story. More general renormalization group studies\cite{BM} have
emphasized the role of a zero-temperature fixed-point ruling the
physics at the transition point. This leads naturally to the presence
of a new exponent $\theta$ that governs hyperscaling violation:
$2-\alpha =\nu (d-\theta)$ where $\nu$ is the correlation length exponent 
and $\alpha$ the specific heat exponent. 
This general scheme is in agreement with the droplet picture of the
transition\cite{Villain,Fisher}. In these theories, there is no
obvious relationship between the random-field exponents as a function
of the dimension and the exponents of the pure Ising system in
dimension $d-\theta$. Stated otherwise, the zero-temperature fixed
point is {\it a priori} unrelated to the ordinary Ising fixed point.
This implies also that there are {\it a priori} three independent
exponents: $\theta$ is a new exponent unrelated to $\nu$ and $\eta$,
the exponent of the decay of spin correlations at the critical point.
However, it has been suggested that dimensional reduction is still
valid\cite{series,Schwartz,Shapir} in a generalized form, i.e.
$\theta =2-\eta$ so that a two-exponent scaling is again valid. 
This means that the exponents of the random-field Ising model
in  dimension $d$ would be those of the pure Ising model in 
dimension $d-\theta$.

In this Letter, we show that the best estimates for the critical
exponents of the random-field Ising model in dimension $d=3$ are in
fact in agreement with the exponents of the pure Ising model
analytically continued in dimension $d^\prime =3 - \theta $. This
analytic continuation is obtained through effective summation methods
of the perturbative $\epsilon = 4-d^\prime $ expansion series for the
critical exponents of the pure Ising system. Such a procedure has
been explored in the past and leads to very stable
estimates\cite{LGZ1,LGZ2}.

The singular part of the free energy in a correlation volume $F_{\rm
sing}$ is governed by the correlation length:
\begin{equation}
F_{\rm sing} \propto \xi^\theta ,
\end{equation}
where the spin correlation length diverges at the critical point
$\xi\propto |T-T_c|^{-\nu}$. This definition leads to a modified
hyperscaling relation:
\begin{equation}
 2-\alpha =\nu (d-\theta).
\end{equation}
Right at the critical temperature, the decay of the connected spin
correlation function can be written:
\begin{equation}
\overline{\langle S_0 S_x\rangle - \langle S_0 \rangle \langle
S_x\rangle} \propto x^{-(d-2+\eta)} .
\end{equation}
For the random field problem, the disconnected susceptibility
has a different kind of scaling:
\begin{equation}
\overline{\langle S_0 \rangle \langle  S_x\rangle} \propto
x^{-(d-4+\overline{\eta})} .
\end{equation}
This defines the two exponents $\eta$ and $\overline{\eta}$. The
general scaling theory of Bray and Moore\cite{BM} gives the following
relation:
\begin{equation}
\theta =2-\overline{\eta}+\eta .
\label{A}
\end{equation}
There is an exact inequality due to Schwartz and
Soffer\cite{SchwartzSoffer}:
\begin{equation}
\overline{\eta} \leq 2 \eta ,
\end{equation}
which in fact appears to be fulfilled as an
equality\cite{Ima,Schwartz,series,RIY}.
The scaling relations between exponents for the pure Ising model in
dimension $d^\prime $ are obtained from those for the random-field
case in dimension $d$ through the relation:
\begin{equation}
d^\prime =d-\theta .
\label{B}
\end{equation}
However, there is no direct evidence that the values of the 
exponents for the
pure Ising model in dimension $d^\prime $ have any relationship with
those for the random-field case in dimension $d$.

To explore the relationship with the critical properties of the pure
Ising model, we have used the estimates of the exponents as a function
of the dimension $d^\prime $ coming from the $\epsilon = 4-d^\prime $
expansion, up to order $\epsilon^5$, of the pure system\cite{LGZ2}.
To extrapolate the series, one needs to use powerful summation
techniques that have been widely studied in the
past\cite{LGZ0,LGZ1,LGZ2}. Let us explain here the
principle of the procedure. If an exponent $E$ is known from the
$\epsilon$-expansion:
\begin{equation}
E(\epsilon )=\sum_k E_k \,\,\epsilon^k ,
\end{equation}
one has to construct the so-called Borel transform $B(\epsilon t)$:
\begin{equation}
E(\epsilon )= \int^\infty_0  dt \, \exp (-t)\, t^\rho \,
B(\epsilon t) ,
\end{equation}
where $\rho$ is an adjustable parameter. The series expansion for the
Borel transform is then given by:
\begin{equation}
B(\epsilon t)=\sum_k E_k\, (\epsilon t)^k / \Gamma (k+\rho +1) ,
\end{equation}
where $\Gamma$ is Euler's function.
The coefficients $E_k$ are known to behave for large $k$ as $E_k \sim
k!a^k k^b$ with some constants $a, b$. The advantage of the Borel
transform is that this translates into a large $k$ behaviour $\sim
a^k k^{b-\rho}$ for its coefficients in the series expansion.
Thus the function $B(\epsilon t)$ is analytic at least in a circle:
the closest singularity of $B(\epsilon t)$ occurs at $\epsilon t
=-1/a$. Assuming that $B(\epsilon t)$ is analytic in the cut plane,
one maps the cut plane onto a circle by use the mapping:
\begin{equation}
\epsilon t={4\over a} {w\over (1-w)^2} = \sum_n \lambda _n w^n.
\end{equation}
The Borel transform is now given through a convergent series in $w$,
which leads to
\begin{equation}
E(\epsilon )=\sum_n B_n \;(\int^\infty_0  dt \, \exp (-t)\, t^\rho
\,  [w(t)]^n \,) .
\end{equation}
Apparent convergence of such an expression is improved by varying
$\rho$ and by introducing other free parameters (see
Ref.\cite{LGZ2}). This leads to stable estimates of the exponent $E$
as a function of $\epsilon$. Errors in the extrapolation procedure
can be estimated by varying the arbitrary parameters (like $\rho$)
that enters the whole procedure. 

Extrapolating down from $d^\prime =4$ to $d^\prime =2$, one finds for
example $\nu =0.99(4)$ for the correlation length exponent and
$\gamma =1.73(6)$ for the susceptibility exponent, in excellent
agreement with the exact values from the Onsager solution $\nu =1$
and $\gamma = 1.75$. As a consequence, it is best to pin the 
exponents
to the exact values in $d^\prime =2$ by writing $E(\epsilon) =
E(d^\prime =2) + (2-\epsilon ) {\tilde E}(\epsilon )$ and to perform
then the above summation method on $\tilde E$. A global check of the
method is provided by the comparison with other methods for $d^\prime
=3$~: this scheme then leads for example to $\nu =0.6310(15)$ while
the standard renormalization group result\cite{LGZ0} from fixed
dimension perturbation series is $\nu =0.6300(15)$, known to be
in agreement with high temperature series and Monte-Carlo results.

Reliable critical exponents $\gamma, \nu , \beta , \eta$ of the pure
system have thus been obtained as a function of the dimension
$d^\prime $. In Figure 1, we plot the results for the
susceptibility exponent $\gamma$. The best value is the central solid
line while the two extremal solid lines define the errors (there is
thus a collapse at $d^\prime =4$ and $d^\prime =2$ of the three
lines). Between $d^\prime =4$ and  $d^\prime =2$ the errors remain
quite small and the value of the exponent does not vary very much.
Below $d^\prime =2$, there is a rapid increase due to the proximity
of the lower critical dimension of the pure problem.

We then consider the random-field Ising model in dimension $d=3$, for
which reliable exponents are now available. The best Monte-Carlo
results\cite{RIY} are obtained using a binary distribution of
random-field. From the corresponding estimates of the optimal run
with ${\overline h}/T=0.35$, one has $\eta = 0.56(3)$ and
$\overline{\eta} = 1.00(6)$, and one then obtains from eq.(\ref{B})
and eq.(\ref{A}): $d^\prime =1.44(10) $. The corresponding value for
$\gamma $ is : $\gamma = 2.3(3)$. These Monte-Carlo results for the
random-field Ising model in dimension $d=3$ appear in Figure 1 as the
solid box : its vertical size is fixed by the estimate of the
exponent $\gamma $ with its error bar and its horizontal size is
fixed by the range of $d^\prime $, which defines the solid box.

In Figure 1, the large intersection between the solid box and the
domain between the extremal solid lines, as well as the good
intersection of the central solid line, clearly shows that $\gamma$
of the three dimensional random problem is in agreement with $\gamma$
of the pure system for $d^\prime =3 - \theta $, and in particular
close to $\gamma$ of the pure system for $d^\prime \approx 1.5$ which
corresponds to the simple guess\cite{RI} that random field 
fluctuations dominate the critical behaviour $F_{sing}\sim 
\xi^{d/2}$: $\theta = 1.5$ in three dimensions.
\begin{figure}
\centering\leavevmode
\epsfxsize=15cm
\epsfbox{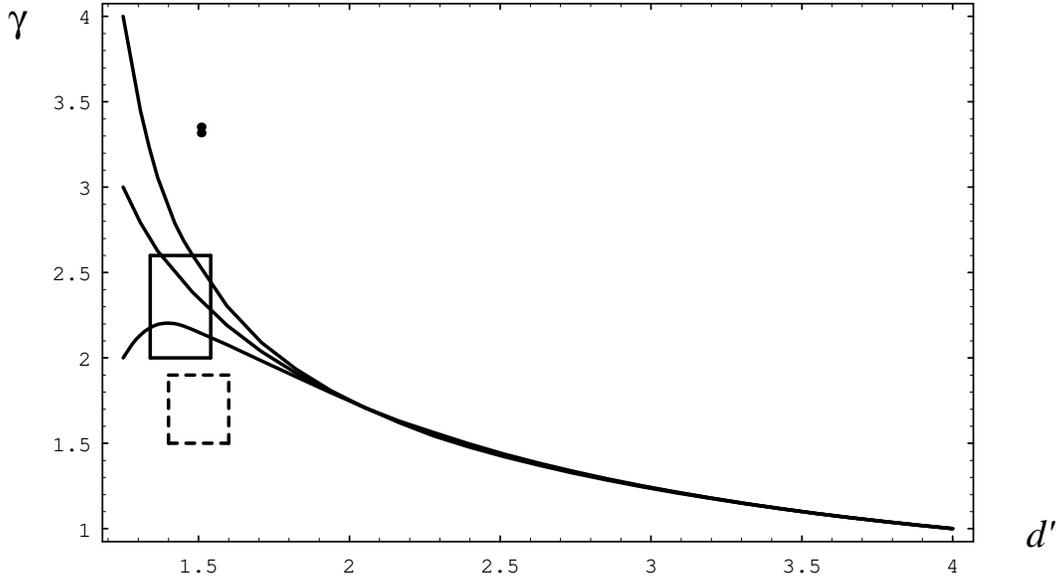}

\caption{Critical exponent $\gamma$ of the magnetic susceptibility
as a function of the dimension  $d^\prime = 3-\theta $. The meaning 
of the various symbols is given in the text.}

\end{figure}
We have also plotted in Figure 1 two less reliable other estimations: 
The dashed box is obtained in the same way as the solid one but
with the results of the Monte-Carlo study of Ref.\cite{RI} using a
Gaussian distribution of the random field, with a slower algorithm
and a less extensive simulation than for the binary distribution. 
The two black dots are estimates from recent Migdal-Kadanoff
renormalization-group studies\cite{RSRG,FH}. Other results
coming from earlier Monte-Carlo simulations\cite{YN} $d^\prime
=1.55(3)$ $\gamma = 1.7(2)$, from real-space renormalization-group
calculations\cite{Dayan} $\gamma = 1.9 \; - \; 2.2$ or from
high-temperature series expansion\cite{series} $\gamma = 2.1(2)$ have
not been drawn on Figure 1 for clarity, since they are well 
compatible with the quoted Monte-Carlo results.
\begin{figure}
\centering\leavevmode
\epsfxsize=15cm
\epsfbox{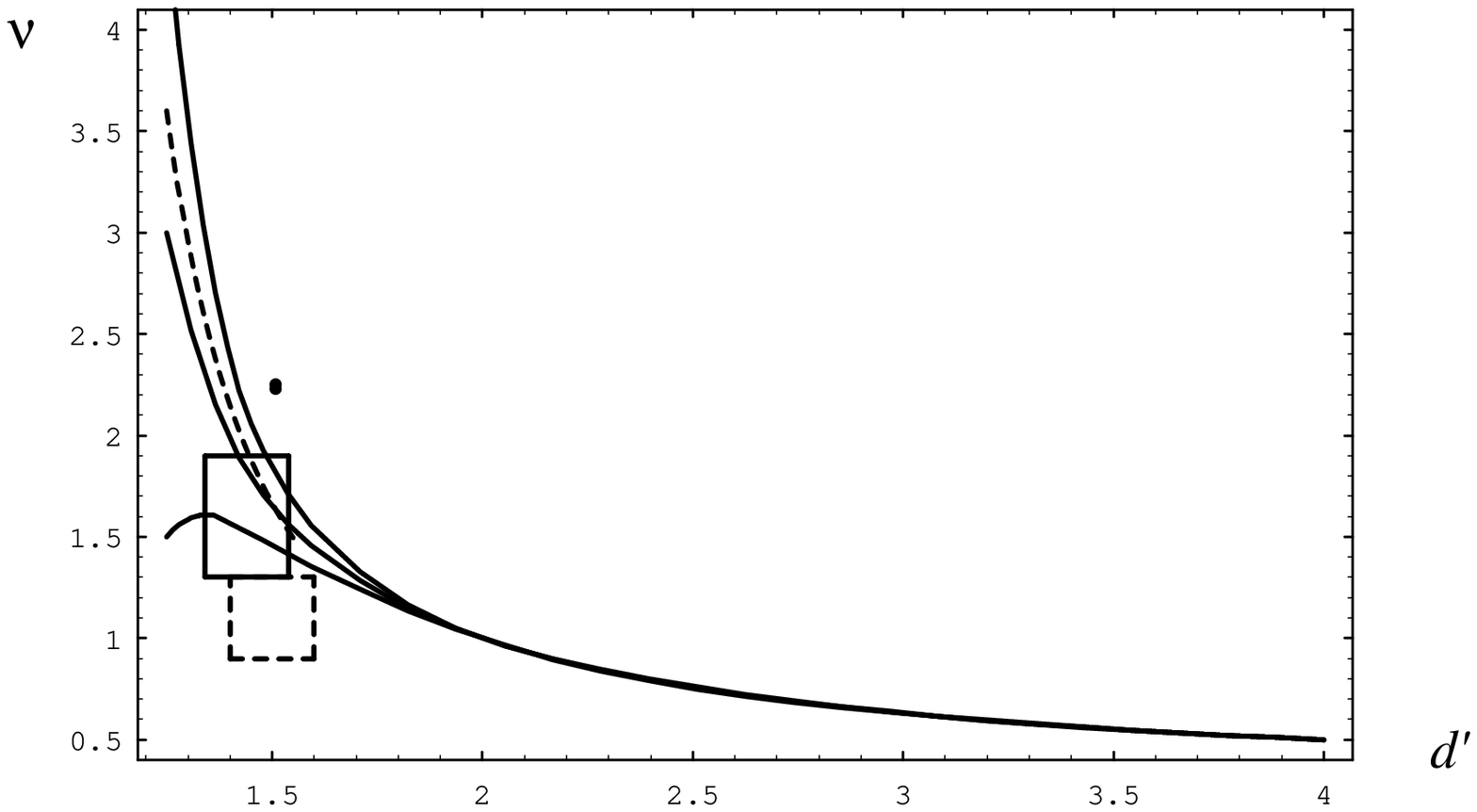}

\caption{Critical exponent $\nu$ of the correlation length
as a function of the dimension  $d^\prime $. 
Same symbols as in Fig. 1}

\end{figure}
In Fig. 2, we plot the results for $\nu$ with the same symbols. The
dashed line is the estimate (up to $d^\prime \approx 1.6$) coming
from the near-planar interface model of Wallace and Zia\cite{WZ}
which has the same critical behaviour as the pure Ising model and can
be expanded in powers of $\epsilon^\prime =d^\prime -1$, giving $\nu$
= $\epsilon^\prime {}^{-1} - 1/2 +  \epsilon^\prime /2$ (We have 
plotted the only non-trivial Pad\'e approximant). Note the
agreement between this estimate and the $\epsilon = 4-d^\prime $
expansion estimates.
\begin{figure}
\centering\leavevmode
\epsfxsize=15cm
\epsfbox{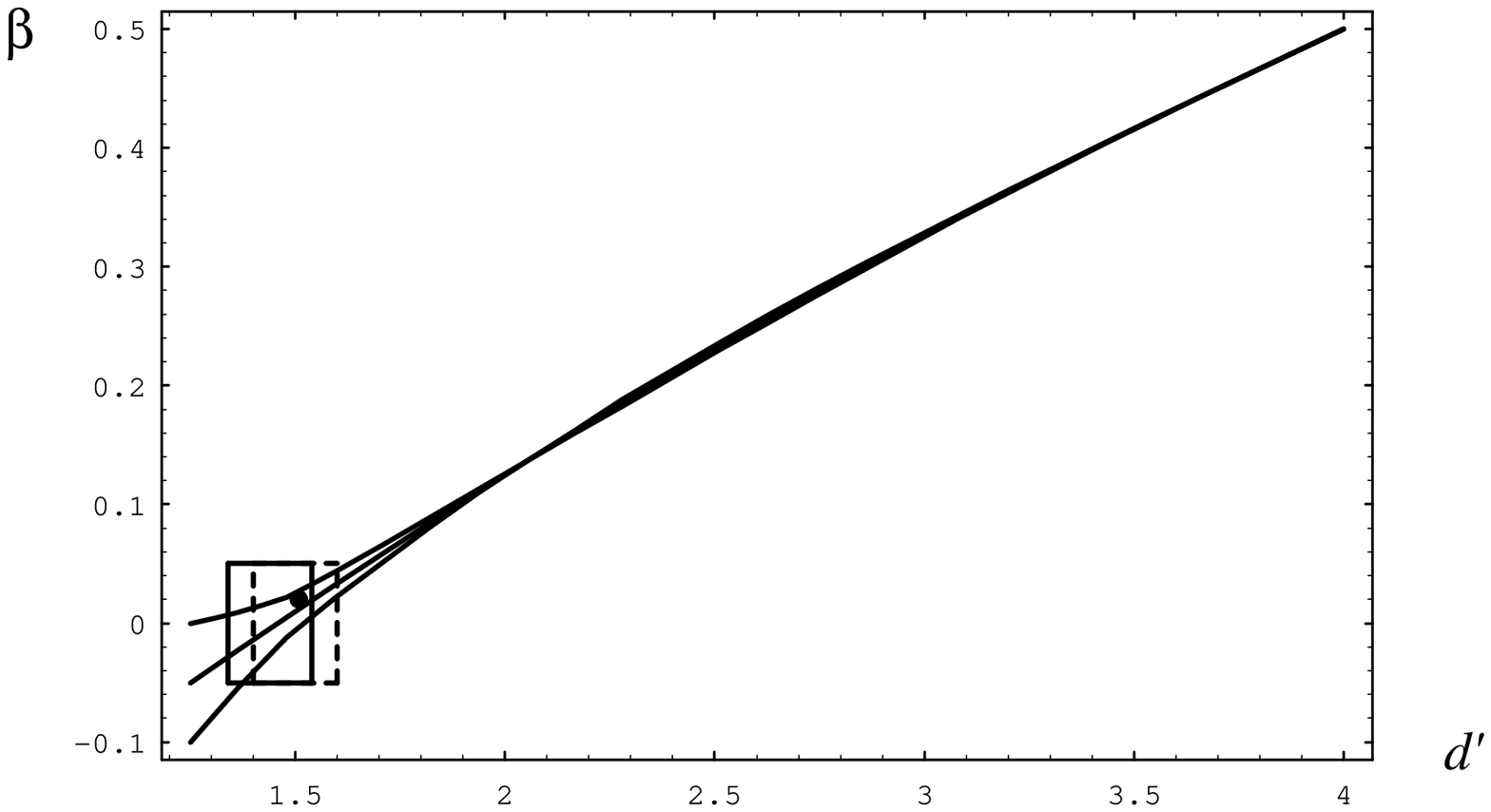}

\caption{Critical exponent $\beta$ of the magnetization
as a function of the dimension  $d^\prime $. Same 
symbols as in Fig. 1}

\end{figure}
In Fig. 3, we plot the order parameter exponent $\beta$. In this
case, it is interesting to note that the Migdal-Kadanoff
calculation\cite{RSRG,FH} shows a non-analytic vanishing of $\beta$ near
$d=2$ and this also happens near $d^\prime =1$ in the droplet model
of Bruce and Wallace\cite{BW} which is an extension of the
near-planar interface model. Accordingly, our resummed results leads
to an extremely small value (or even negative, the non-analytic
behavior cannot be possibly reproduced by our approximants) of
$\beta$ for the corresponding dimension.

\begin{figure}
\centering\leavevmode
\epsfxsize=15cm
\epsfbox{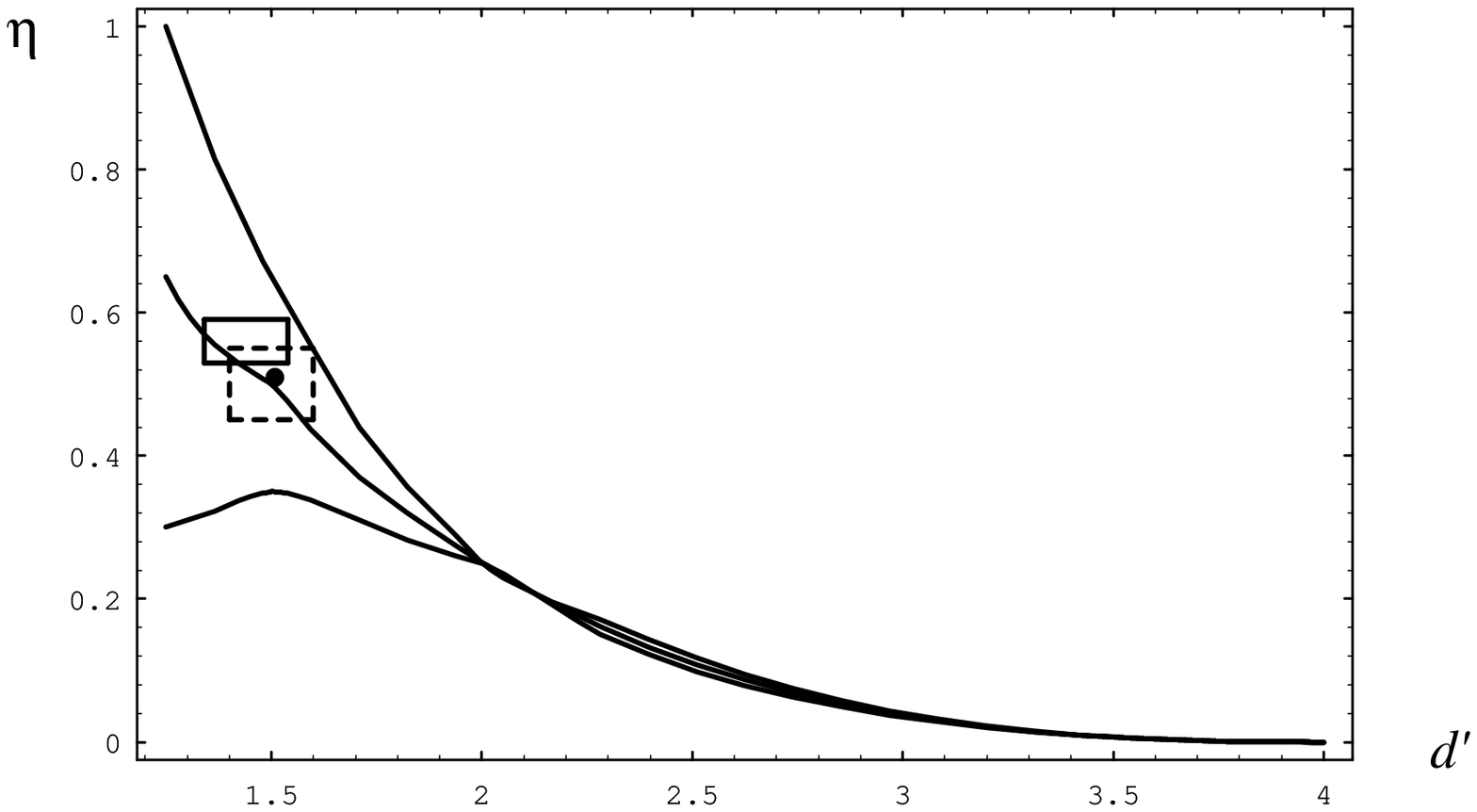}

\caption{Critical exponent $\eta$ of the spin correlations
as a function of the dimension  $d^\prime $. 
Same symbols as in Fig. 1}

\end{figure}
Finally, in Fig. 4, we plot $\eta$, again with the same symbols.
In this case, as well as for $\beta$, the Monte-Carlo results are in
good agreement with the Migdal-Kadanoff values.

The overall remarkable agreement clearly seen in our Figures between
the results for the three dimensional random-field Ising model and
those for the pure case for the corresponding dimension shows that
the exponents of the random-field Ising model in three dimensions are
in very good agreement with those of the pure Ising system in
dimension $d^\prime  =3-\theta$, and in particular very close to
those of the pure system in dimension $\approx 1.5$ which corresponds
to the {\it a priori} guess\cite{RI} $\theta = 1.5$ in three
dimensions.

We have thus obtained evidence for the generalized dimensional 
reduction for the random-field Ising model: the values of
the exponents for the random model in dimension $d$ are
in fact those for the pure case in dimension $d^\prime =d-\theta$.

\acknowledgments

We would like to thank J. Y. Fortin and P. C. W. Holdsworth for
useful discussions.


\end{document}